\documentclass[aps,prb,twocolumn,superscriptaddress,showpacs,floatfix]{revtex4}

\usepackage{graphicx}
\graphicspath{{figs/}}
\begin{document}
\title{Thermal expansion in single-walled carbon nanotubes and graphene: nonequilibrium Green's function approach}
\author{Jin-Wu~Jiang}
    \affiliation{Department of Physics and Centre for Computational Science and Engineering,
             National University of Singapore, Singapore 117542, Republic of Singapore }
\author{Jian-Sheng~Wang}
    \affiliation{Department of Physics and Centre for Computational Science and Engineering,
                 National University of Singapore, Singapore 117542, Republic of Singapore }
\author{Baowen~Li}
        \affiliation{Department of Physics and Centre for Computational Science and Engineering,
                     National University of Singapore, Singapore 117542, Republic of Singapore }
        \affiliation{NUS Graduate School for Integrative Sciences and Engineering,
                     Singapore 117456, Republic of Singapore}
\date{\today}
\begin{abstract}
The nonequilibrium Green's function method is applied to investigate
the coefficient of thermal expansion (CTE) in single-walled carbon
nanotubes (SWCNT) and graphene. It is found that atoms expand about
1$\%$  from equilibrium positions even at $T=0$ K, resulting from the
interplay between quantum zero-point motion and nonlinear
interaction. The CTE in SWCNT of different sizes is studied and
analyzed in terms of the competition between various vibration
modes. As a result of this competition, the axial CTE is positive in
the whole temperature range, while the radial CTE is negative at low
temperatures. In graphene, the CTE is very sensitive to the
substrate. Without substrate, CTE has large negative region at low
temperatures and very small value at high temperature limit, and the
value of CTE at 300 K is $-6\times 10^{-6}$ K$^{-1}$ which is
very close to a recent experimental result, $-7\times 10^{-6}$
K$^{-1}$ [Nat. Nanotechnol. \textbf{10}, 1038 (2009)]. A very weak
substrate interaction (about 0.06$\%$ of the in-plane interaction)
can largely reduce the negative CTE region and greatly enhance the
value of CTE. If the substrate interaction is strong enough, the CTE
will be positive in whole temperature range and the saturate value
at high temperatures reaches $2.0\times 10^{-5}$ K$^{-1}$.
\end{abstract}

\pacs{65.80.+n, 05.70.-a, 61.46.-w, 62.23.Kn} \maketitle

\section{introduction}
Single-walled carbon nanotubes (SWCNT) and graphene are two kinds of
novel carbon based materials with lots of intriguing electronic and
mechanical properties.\cite{DaiHJ, Dresselhaus, Novoselov, Avouris,
Neto, Geim} They have many potential applications in nano-devices,
where the thermal property plays a big role. The coefficient of
thermal expansion (CTE) is one of the most important nonlinear
thermal properties. Very recently, the CTE of graphene is measured
to be $-7\times 10^{-6}$ K$^{-1}$ at $T=300$ K.\cite{BaoW} Also the
CTE of multi-walled carbon nanotubes\cite{Ruoff, Bandow, Maniwa1}
and nanotube bundles\cite{Yosida, Maniwa2, Dolbin, Dolbin2009} are
measured in experiments. Theoretically,  there are several
approaches to calculate CTE such as molecular
dynamics,\cite{Raravikar, Schelling, Cao, KwonYK} lattice dynamics
calculations,\cite{Schelling, Barron, Fabian, Broido, Ward} the
molecular structural mechanics method\cite{LiC}, analytical
method\cite{Jiang}, and \textit{ab initio} density-functional theory
and density-functional pertubation theory calculations.\cite{Mounet,
Bonini} However, discrepancies exist in different approaches. In
Ref.~\onlinecite{Jiang, Schelling, KwonYK}, the CTE is predicted to
be negative in low temperature region while positive at high
temperatures. On the contrary, Li and Chou obtain positive value for
CTE in whole temperature range.\cite{LiC} In
Ref.~\onlinecite{Mounet}, CTE for graphite is negative at low
temperatures and positive at high temperatures, and CTE for graphene
is negative in whole temperature range. In Refs.~\onlinecite{Jiang,
KwonYK}, the axial CTE increases from negative to positive with
increasing temperature; while the minimum value of axial CTE is
quite different: $-1\times 10^{-6}$ K$^{-1}$ or $-5\times 10^{-5}$
K$^{-1}$ and this value is achieved at different temperatures: 300 K
or 500 K in Ref.~\onlinecite{Jiang} and Ref.~\onlinecite{KwonYK}
respectively. In the high temperature region, the value for CTE is
also quite different from one to another, typically about $2.0\times
10^{-6}$ K$^{-1}$ in Refs.~\onlinecite{LiC,  Schelling}, $4.0\times
10^{-6}$ K$^{-1}$ in Ref.~\onlinecite{Jiang} or more than
$5\times10^{-6}$ K$^{-1}$ in Ref.~\onlinecite{KwonYK}. In
Ref.~\onlinecite{LiC} the CTE shows only a slight dependence on the
length or diameter of the SWCNT. While Ref.~\onlinecite{Jiang}
predicts the obvious decrease of CTE with increasing diameter. To
the best of our knowledge,  the limited existing theoretical works
are diverse.

In this paper, we investigate the CTE in SWCNT and graphene sheets
by the nonequilibrium Green's function (GF) approach, which includes
contributions from all phonon modes and takes into account  the
quantum effect. Our study shows that even at $T=0$ K the averaged
position of each atom deviates about 1$\%$ from its equilibrium
position. This is the result of the combined contribution from
quantum zero-point motion and nonlinear interaction. For the CTE in
SWCNT, we study the competition between lateral bending vibration
(negative effect), the radial breathing mode (positive effect), and
the longitudinal optical phonon mode (positive effect). As a result,
the axial CTE is positive in the whole temperature range; while the
radial CTE has negative value at low temperatures and the radial CTE
is in general one order of magnitude smaller than the axial CTE at
high temperatures. Our results show that the axial CTE is not very
sensitive to the size of SWCNT. Yet for radial CTE, the thinner
SWCNT (with higher aspect ratio length/diameter, $L/d$) has more
tense bending vibration, thus has smaller value. For graphene, we
demonstrate that the value of CTE is very sensitive to the
interaction $\gamma$ between the substrate and graphene. If there is
no substrate, graphene has negative CTE at low temperatures and
generally the CTE is very small at high temperatures, and the value
of CTE at $T=300$ K from our calculation is $\alpha=-6\times
10^{-6}$ K$^{-1}$, which agrees with the experimental one. When the
interaction is nonzero, even a very small value ($\gamma=0.001$
eV/(\AA$^{2}$u)) will largely shrink the negative CTE region, and
greatly enhance the value of CTE at high temperatures. If $\gamma$
is larger than 0.1 eV/(\AA$^{2}$u), the CTE is positive in the whole
temperature range and saturates at value of $2.0\times 10^{-5}$
K$^{-1}$ at high temperature.

The paper is organized as follows. In Sec.~II, we present the
detailed derivations in the GF method. Sec.~III is devoted to the
main results  and relevant discussions. The conclusion is in
Sec.~IV.

\section{Green's function method}
The definition and notations of GF used in this paper are from
Ref.~\onlinecite{WangJS2007, WangJS2008, Zeng}, where relationships
between different versions of GF can also be found. In the present
investigation of CTE, two types of GF are required :
\begin{eqnarray}
G_{j}(\tau) & = & -\frac{i}{\hbar}\langle T_{\tau}u_{j}^{H}(\tau)\rangle,
\label{eq_gone}\\
G_{jk}(\tau,\tau') & = & -\frac{i}{\hbar}\langle T_{\tau}u_{j}^{H}(\tau)u_{k}^{H}(\tau')\rangle,
\end{eqnarray}
where $u_{j}^{H}(\tau)$ is the vibrational displacement of atom $j$ in Heisenberg picture, multiplied by the square root of mass of the atom.
$T_{\tau}$ is the contour-order operator on the Keldysh contour and $\tau$ is the time on the contour. The contour-order operator in $G_{j}(\tau)$ is necessary in the further perturbation expansion of it. In the harmonic system without nonlinear interaction, all atoms vibrate around the equilibrium position linearly, so $G_{j}^{0}(\tau)=0$ and the two-point harmonic contour ordered GF $G_{jk}^{0}(\tau,\tau')$ can be exactly calculated:
\begin{eqnarray}
G_{jk}^{0}(\tau,\tau') & = & -\frac{i}{\hbar}\langle T_{\tau}u_{j}^{H}(\tau)u_{k}^{H}(\tau')\rangle_{0},
\end{eqnarray}
where the subscript 0 indicates a harmonic system.

These two GFs are defined in the Heisenberg picture, however the use
of interaction picture is more advantageous in the study of systems
with nonlinear interaction. The $G_{j}(\tau)$ in the interaction
picture can be obtained by using the general picture
transformations:
\begin{eqnarray}
G_{j}(\tau)  =  -\frac{i}{\hbar}\langle T_{\tau}S(-\infty,-\infty)u_{j}^{I}(\tau)\rangle,
\end{eqnarray}
where $u_{j}^{I}$ is the vibrational displacement
in the interaction picture.
$S$ is the scattering matrix. The nonlinear interaction used in the
present paper takes the form:
\begin{eqnarray}
H_{n} & = & \sum_{lmn}\frac{k_{lmn}}{3}u_{l}u_{m}u_{n}+\sum_{opqr}\frac{k_{opqr}}{4}u_{o}u_{p}u_{q}u_{r},
\label{eq_nonlinear}
\end{eqnarray}
where $k_{lmn}$ and $k_{opqr}$ are
interaction constants for the
cubic and quartic interactions. These nonlinear interactions are
generated from the Brenner potential\cite{Brenner} by finite
difference method.

\begin{figure}[htpb]
  \begin{center}
    \scalebox{1.2}[1.2]{\includegraphics[width=7cm]{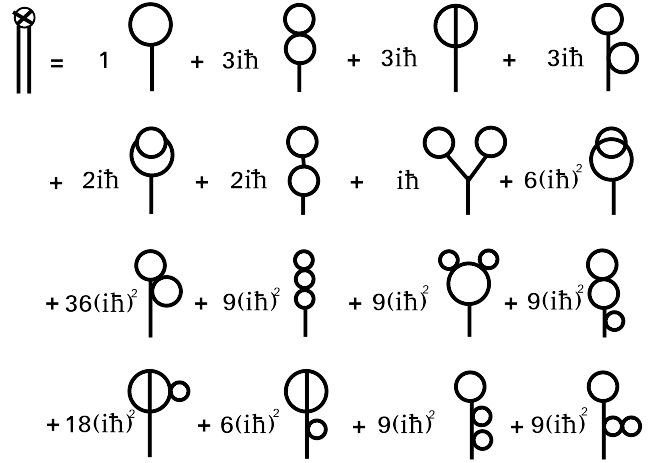}}
  \end{center}
  \caption{Feynman diagrams for $G_{j}(\tau)$, with the cubic and quartic
  nonlinear interactions up to the second order of $\hbar$.}
  \label{fig_feynman}
\end{figure}
By applying Wick's theorem, $G_{j}(\tau)$ can be expanded in terms
of $H_{n}$. Fig.~\ref{fig_feynman} gives the Feynman diagrams for
$G_{j}(\tau)$ up to the second order of $\hbar$. For thermal
expansion, the first diagram in this figure is most important, so we
mainly consider this diagram in the following context. The
contribution of this diagram to $G_{j}(\tau)$ is:
\begin{eqnarray}
G_{j}(\tau) & = & \sum_{lmn}\int d\tau_{1}d\tau_{1}'d\tau_{1}''T_{lmn}(\tau_{1},\tau_{1}',\tau_{1}'')G_{lm}^{0}(\tau_{1},\tau_{1}')\nonumber\\
&   &G_{nj}^{0}(\tau_{1}'',\tau),
\end{eqnarray}
where we have introduced $T_{lmn}(\tau_{1},\tau_{1}',\tau_{1}'')=k_{lmn}\delta(\tau_{1}',\tau_{1})\delta(\tau_{1}'',\tau_{1})$ for convenience. The contour ordered GF can be transformed into time domain by using the corresponding rule:\cite{WangJS2008} $\tau \rightarrow (t, \sigma)$, where the branch index $\sigma=\pm$ is introduced such that $\tau=t+i\epsilon \sigma$ is on the upper ($\sigma=+1$) or lower ($\sigma=-1$) contour branches. $\epsilon$ is a small positive number. We find that $G_{j}^{\sigma}(t)$ is independent of the branch index $\sigma$ and time $t$. This can be understood in the sense that the system after thermal expansion is in thermal equilibrium state, and the thermal expansion effect does not depend on the direction of the contour. For simplicity, we use $G_{j}$ to represent the one point GF $G_{j}^{\sigma}(t)$ in the following. After some derivations and using the Langreth rules (see appendix), we obtain the final expression for $G_{j}$:
\begin{eqnarray}
G_{j}  = \sum_{lmn}k_{lmn}G_{lm}^{>}(0)\tilde{G}_{nj}^{r}[0],
\label{eq_gj}
\end{eqnarray}
where $G^{>}$ and $\tilde{G}^{r}$ are the greater and retarded GF. They can be calculated in the eigen space without doing any integration as given in the appendix. For simplicity we have omitted the superscript 0, which indicates a harmonic system. To avoid confusion, we use $G(t)$ in time domain and $\tilde{G}[\omega]$ in frequency domain. We note that the information of temperature $T$ is carried by $G^{>}$. Then the averaged vibrational displacement for atom $j$ can be obtained from Eq.~(\ref{eq_gone}). From now on, the symbol $u_{j}$ will denote the averaged vibrational displacement for atom $j$. This is different from its original meaning in Eq.~(\ref{eq_nonlinear}) where no average is taken. The CTE at temperature $T$ is calculated from:
\begin{eqnarray}
\alpha_{j} = \frac{i\hbar}{M}\times\frac{1}{x_{j}}\times \frac{dG_{j}}{dT},
\label{eq_cte}
\end{eqnarray}
where $M$ is the mass of atom. $x_{j}$ is the position of atom $j$,
and $x_{j}=0$ at the common boundary of the fixed and the other
regions. Using this equation, the CTE is obtained from the
information of a single atom $j$. To obtain more accurate value for
CTE, one needs to average $\alpha_{j}$ over the atoms.

\section{Results and discussions}
\begin{figure}[htpb]
  \begin{center}
    \scalebox{1.0}[1.0]{\includegraphics[width=7cm]{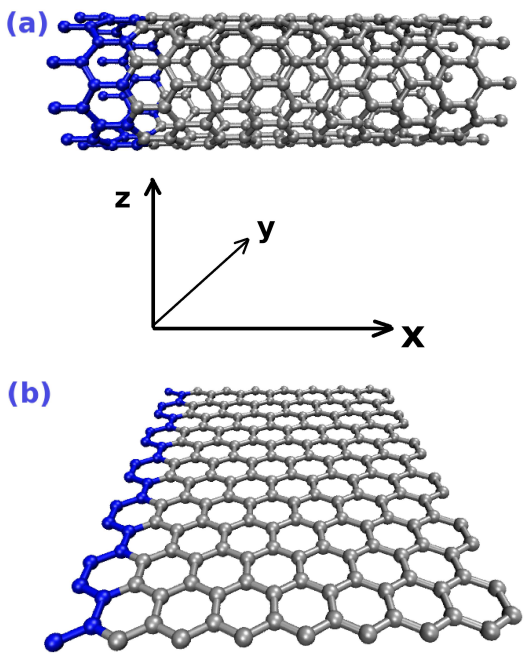}}
  \end{center}
  \caption{(Color online)
 Configurations for different systems: (a) Single-walled carbon
nanotube; (b) Graphene sheet. The left boundary parts (blue online)
are fixed.}
  \label{fig_cfg}
\end{figure}
In our calculations, the system is relaxed using the Brenner's
potential. Then the dynamical matrix and nonlinear force constants
are obtained by finite difference method. The dynamical matrix is
diagonalized to obtain the eigen frequencies and eigen vectors.
Finally, the eigen frequencies, eigen vectors and nonlinear force
constants are used to calculate the GF following the
formula in the appendix. Generally, the relevant anharmonic value
from Brenner's potential is about
$|k_{lmn}/k_{ij}|\approx0.8${\AA$^{-1}$}, where $k_{ij}$ is one
element in the dynamical matrix.

Fig.~\ref{fig_cfg} is the configurations for SWCNT and graphene in
our calculation. For both systems, the smaller balls (blue online)
on the left boundaries are fixed atoms while the right boundaries
are free. This boundary condition corresponds to thermal expansion
under zero pressure. The $x$ axis in SWCNT is along its axial. In
graphene, the $x$ axis is along the in-plane horizontal direction,
$y$ in the vertical direction and $z$ in the perpendicular
direction. Periodic boundary condition is applied in the $y$
direction in graphene.

\begin{figure}[htpb]
  \begin{center}
    \scalebox{1.0}[1.0]{\includegraphics[width=7cm]{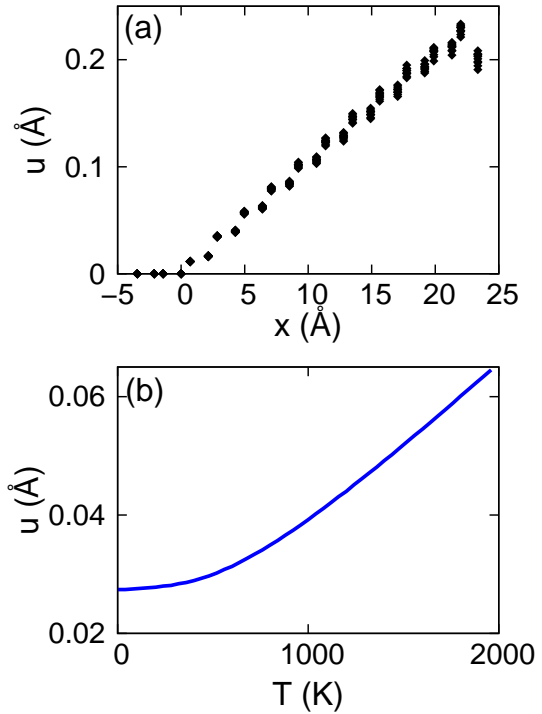}}
  \end{center}
  \caption{The averaged vibrational displacement $u$ of SWCNT. (a). $u$ v.s. $x$ at $T=1000$ K. (b).
$u$ v.s. $T$ for the atom located at $x=3$~{\AA}.}
  \label{fig_vibdisp}
\end{figure}
We firstly study the axial thermal expansion in SWCNT.
Fig.~\ref{fig_vibdisp}~(a) shows the averaged vibrational
displacement in the $x$ direction for different atoms, i.e., $u(x)$.
The horizontal axis is the position of each atom along $x$ axis.
Generally, $u(x)$ increases with increasing $x$ which reveals the
fact that the thermal expansion is homogenous along $x$ axis. On the
two boundaries, there are some boundary effects due to the different
local environments from atoms in the center region.
Fig.~\ref{fig_vibdisp}~(b) shows the temperature dependence for $u$
of one of the atoms in the center region. At high temperatures,
$u\propto T$ is the classical behavior. With temperature decreasing,
quantum effect appears, and around $T=800$ K, the curve deviates
distinctly from the classical behavior. This implies that the Debye
temperature in the SWCNT is typically about 800 K. The quantum
effect dominates the low temperature region with $T < 150$ K. In
this temperature region, $u$ keeps a nonzero constant, resulting
directly from the combination effect of the thermal zero-point
motion and nonlinear interaction in SWCNT. It means that even at
$T=0$ K, atoms still vibrate around equilibrium positions, and as a
result of the nonlinear interaction, their averaged vibrational
displacements is about 1$\%$ of lattice constant from equilibrium
positions. This value is not small because the interaction in
covalent bonding carbon system is very strong, so the frequencies of
the optical phonon modes are high, thus the zero-point motion energy
is also high. We expect this temperature dependent curve for $u$ can
be confirmed by x-ray or other optical experiments.

\begin{figure}[htpb]
  \begin{center}
    \scalebox{1.1}[1.1]{\includegraphics[width=7cm]{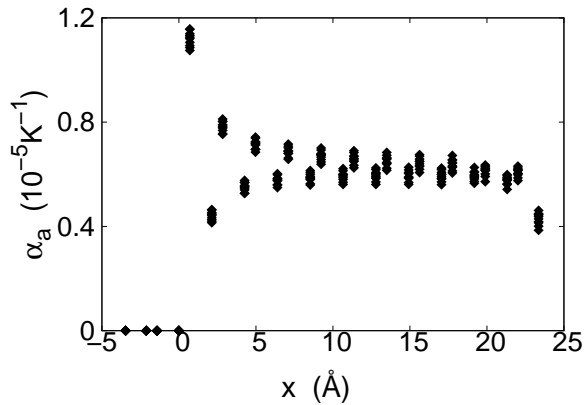}}
  \end{center}
  \caption{(Color online) CTE of SWCNT calculated
  from the thermal expansion of atoms at different location.}
  \label{fig_cte_a_x}
\end{figure}
\begin{figure}[htpb]
  \begin{center}
    \scalebox{1.0}[1.0]{\includegraphics[width=7cm]{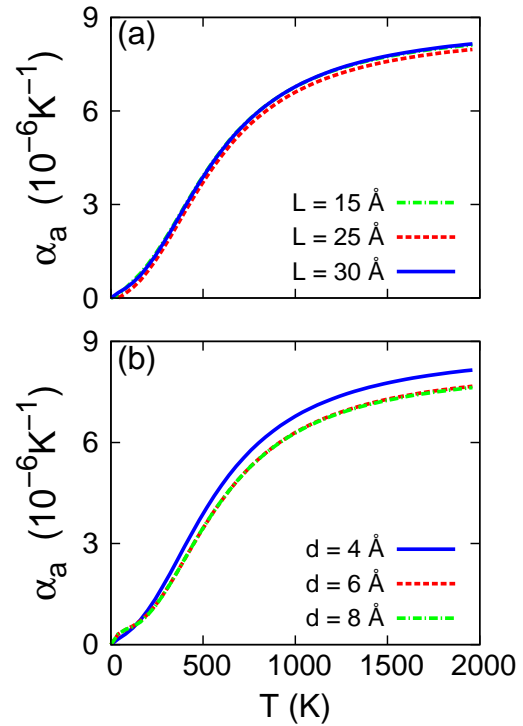}}
  \end{center}
  \caption{(Color online) Axial CTE v.s. temperature in SWCNT with: (a). diameter $d=4$~{\AA}, and different length $L$; (b). length $L=30$~{\AA}, and different diameter $d$.}
  \label{fig_cte_a_T}
\end{figure}
The axial CTE ($\alpha_{a}$) of SWCNT can be calculated by using
$u_{j}$ of each atom from Eq.~(\ref{eq_cte}). Fig.~\ref{fig_cte_a_x}
shows the value of axial CTE calculated from different atoms at
$T=1000$ K. The horizontal axis is the $x$ position for each atom.
SWCNT in this figure is zigzag (5, 0), with length $L=30$~{\AA}. Due
to boundary effects on the left and right ends, values for CTE in
these two regions deviate obviously from that in the center region.
So we drop data on two boundaries and average over other data in the
center region to obtain the value of CTE.

Fig.~\ref{fig_cte_a_T} shows the temperature dependence for axial
CTE in zigzag SWCNT of different sizes. They are similar to the
results of armchair SWCNT. It shows that the length and diameter
have small effects on axial CTE, and the value of axial CTE is
positive in the whole temperature region. This is consistent with
previous results in Ref.~\onlinecite{LiC}, but different from that
in Ref.~\onlinecite{Jiang}, where the axial CTE depends on diameter
sensitively and is negative at low temperature. In general, the
value of CTE increases quickly in low temperature region, and
reaches the classical limit at high temperatures. We point out that
at $T=800$ K, this value still changes a lot with changing
temperature. This temperature (800 K) is a bit smaller than the
value of Debye temperature in SWCNT or graphene, which is above 1000
K.\cite{Benedict, Tohei, Falkovsky} This result can be understood in
the sense that the optical phonon modes play an important role in
the calculation of the Debye temperature.\cite{Tohei} While in the
study of the thermal expansion effect, the optical phonon modes are
not very important due to their high frequency. So the critical
temperature here is lower than Debye temperature.

\begin{figure}[htpb]
  \begin{center}
    \scalebox{1.0}[1.0]{\includegraphics[width=7cm]{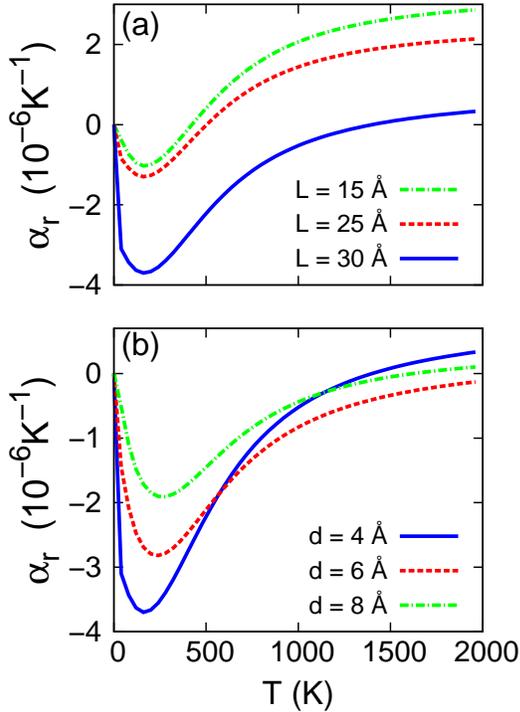}}
  \end{center}
  \caption{(Color online) Radial CTE of SWCNT with (a). $d=4$~{\AA} and different length; (b). $L=30$~{\AA} and different diameter.}
  \label{fig_cte_r}
\end{figure}
\begin{figure}[htpb]
  \begin{center}
    \scalebox{1.0}[1.0]{\includegraphics[width=7cm]{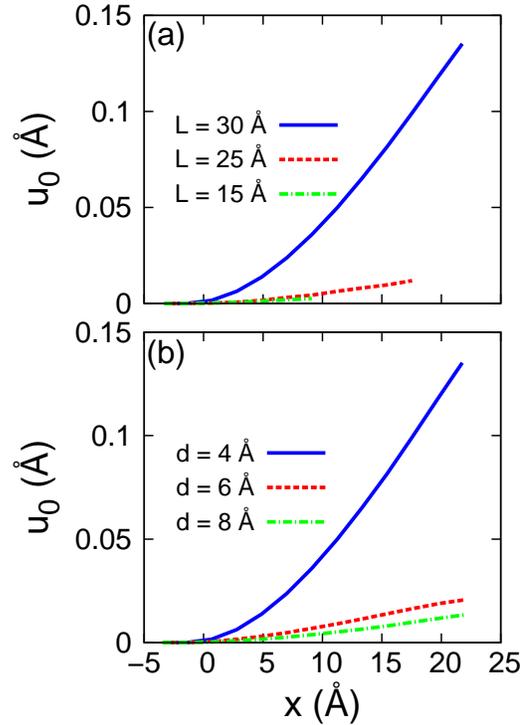}}
  \end{center}
  \caption{(Color online) The averaged position of axial line at $T=1000$ K in zigzag SWCNT with (a). $d=4$~{\AA} and different length; (b). $L=30$~{\AA} and different diameter.}
  \label{fig_midline}
\end{figure}

Fig.~\ref{fig_cte_r} is the temperature dependence for the radial
CTE ($\alpha_{r}$) of different sized SWCNT. The radial CTE is
obtained from the changing of diameter with changing temperature.
Compared with the axial CTE, there are abundant behaviors here. (1).
Firstly, we can see $\alpha_{r} < 0$ in low-temperature region. The
origin of this phenomenon is the bending vibration in this rod-like
SWCNT system.\cite{Krishnan} Fig.~\ref{fig_midline} shows the
position for midline in SWCNT along the $x$ axis at $T=1000$ K. This
figure indicates the bending vibration directly. This bending
vibration has a very low energy thus is very important at low
temperatures. It leads to contraction in radial direction in the
SWCNT. As a result, the radial CTE is negative in low temperature
region where other high energy vibration modes are not excited. (2).
A general behavior in both Fig.~\ref{fig_cte_r}~(a) and (b) is that
the thinner SWCNT (with higher aspect ratio length/diameter, $L/d$)
has smaller value of $\alpha_{r}$ and larger negative CTE region.
This is mainly because the bending vibration in thinner SWCNT is
more intense (see Fig.~\ref{fig_midline}). For SWCNT with different
diameters shown in Fig.~\ref{fig_cte_r}~(b), there is one more
reason. In SWCNT, another important vibrational mode in the radial
direction is the radial breathing mode. This mode can make the SWCNT
expand in the radial direction. The frequency of the radial
breathing mode is inverse proportional to diameter. So in SWCNT, the
larger the diameter, the smaller the frequency of radial breathing
mode. Therefore, more positive contribution at low temperatures will
lead to larger value of $\alpha_{r}$. The competition between the
bending vibration and the radial breathing vibration gives a valley
in all curves of $\alpha_{r}$ in Fig.~\ref{fig_cte_r}. (3). The
value of $\alpha_{r}$ is typically one order of magnitude smaller than
$\alpha_{a}$. Because along the axial of SWCNT there are some high
frequency optical phonon modes, which lead to strong expansion in
the axial direction. Another reason is that along the circular
direction, SWCNT forms a closed configuration, which makes it more
difficult to expand in the radial direction. While in the axial
direction, the right side is open and expansion along this direction
is easier. In Fig.~\ref{fig_cte_r}~(a), the CTE does not show
convergence behavior with increasing $L$. This is mainly due to very
serious bending vibration in long SWCNT. As a result of this strong
bending movement, it will be difficult for our method to calculate
the radial CTE accurately.

\begin{figure}[htpb]
  \begin{center}
    \scalebox{1.1}[1.1]{\includegraphics[width=7cm]{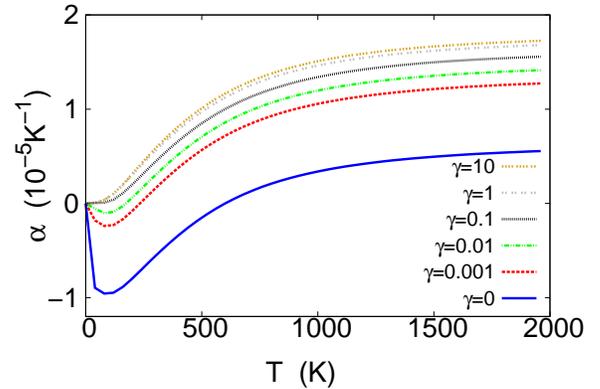}}
  \end{center}
  \caption{(Color online) CTE v.s. temperature
  for graphene sheet with various
 strength of interaction between substrate and graphene. The
interaction $\gamma$ is from 0 to 10 eV/(\AA$^{2}u)$.}
  \label{fig_cte_graphene}
\end{figure}
Now we turn to the thermal expansion in graphene. Graphene is easily
bent due to its two-dimensional elastic sheet configuration. So in
the experiment it is very difficult to hang a graphene sheet by just
holding it on one side (e.g. left side as shown in
Fig.~\ref{fig_cfg}~(b)). Furthermore, to study the thermal expansion
effect, one can not suspend the graphene sheet by fixing both left
and right sides. The only reasonable choice is to put the graphene
sample on a substrate. In this situation, the substrate can support
the graphene stably. But in the mean time, interactions between the
substrate and graphene will affect the bending vibration seriously,
thus influences the measured value for CTE in graphene sheet. In the
following we focus on this substrate effect on the CTE in graphene
sheet.

The interaction between substrate and carbon atoms in graphene can
be described by the on-site potential in the $z$
direction\cite{Aizawa}: $V=(\gamma /2) u_{z}^{2}$ with $\gamma$ as
the interaction force constant in unit of eV/(\AA$^{2}$u). $u_{z}$
is the vibrational displacement in the $z$ direction.
Fig.~\ref{fig_cte_graphene} is the CTE of graphene with 204 atoms
for different values of $\gamma$. CTE is very sensitive to $\gamma$.
When there is no substrate interaction the bending vibration is very
strong (which has negative effect on CTE), so the CTE is very small
in the high temperature limit, and a large negative CTE region
occurs at low temperatures. At $T=300$ K, $\alpha=-6\times 10^{-6}$
K$^{-1}$. This value agrees with the very recent experimental result
of $-7\times 10^{-6}$ K$^{-1}$.\cite{BaoW} However, when the
substrate interaction is introduced, even a very small value of
$\gamma=0.001$, can significantly enhance the value of CTE.
Furthermore, the negative CTE region considerably shrinks. After
further increasing interaction, the value of CTE is further
increased; the negative CTE region is further reduced and disappears
after $\gamma=0.1$. A saturate value for CTE is obtained after
$\gamma=1.0$, where the bending vibration has been suppressed
completely by the substrate interaction. We should pointout that the
value of $\gamma=0.001$ is actually only about 0.06$\%$ of the
interaction between carbon atoms in the two-dimensional graphene
sheet\cite{SaitoR} which is more than 1.67 eV/(\AA$^{2}$u). The
practical experimental substrate interaction might be larger than
this value. So the measured value of CTE will be higher than the
curve with $\gamma=0.001$ in Fig.~\ref{fig_cte_graphene}. If the
experimental substrate interaction is larger than $\gamma=0.1$, no
negative CTE should be observed. By the way, if we compare the
saturated value at high temperatures with that in the SWCNT as shown
in Fig.~\ref{fig_cte_a_T}, we find that the value of CTE in graphene
is much larger than SWCNT. This fact implies that the bending
vibration in the SWCNT is also very important for the axial CTE. It
considerably reduce the value of axial CTE without causing negative
axial CTE in SWCNT.

\section{conclusion}
In conclusion, the GF method has been applied to
investigate the CTE in SWCNT and graphene. The effect of quantum
zero-point motion and nonlinear interaction at $T=0$ K has been
observed clearly. The axial CTE in SWCNT is positive in whole
temperature range, while the radial CTE can have negative value at
low temperatures. It shows that thinner SWCNT has smaller radial CTE,
resulting from the competition between bending vibration and radial
breathing vibration. Our calculation displays that the CTE in
graphene sheet is very sensitive to the interaction between
substrate and graphene, which will greatly enhance the CTE in
graphene sheet.

\section{Acknowledgements}
We thank E. Cuansing and Yong Xu for helpful discussions. The work
is supported by a Faculty Research Grant of R-144-000-173-101/112 and R-144-000-257-112 of
National University of Singapore (NUS), and Grant R-144-000-203-112
from Ministry of Education of Republic of Singapore, and Grant
R-144-000-222-646 from NUS.

\appendix
\section{Langreth rules}
To arrive at Eq.~(\ref{eq_gj}), relations between different types of Green's functions are in need.\cite{WangJS2008} The following Langreth rules\cite{Langreth, Haug} are used to obtain these relations.

For the time invariant contour integral
\begin{eqnarray}
A(\tau,\tau')=\int B(\tau,\tau'')C(\tau'',\tau')d\tau'',
\end{eqnarray}
its Fourier transform has following relations:
\begin{eqnarray}
&&A^{<}[\omega] = B^{r}[\omega]C^{<}[\omega] + B^{<}[\omega]C^{a}[\omega],\nonumber\\
&&A^{>}[\omega] = B^{r}[\omega]C^{>}[\omega] + B^{>}[\omega]C^{a}[\omega],\nonumber\\
&&A^{r}[\omega] = B^{r}[\omega]C^{r}[\omega],\nonumber\\
&&A^{a}[\omega] = B^{a}[\omega]C^{a}[\omega].
\end{eqnarray}

\section{Green's function in eigen space}
The retarded GF $G^{r}[\omega]$ can be obtained straightforwardly, since it has the expression as\cite{WangJS2008}:
\begin{eqnarray}
G_{0}^{r}[\omega] & = & [(\omega+i\eta)^{2}I-K]^{-1},
\end{eqnarray}
where $K$ is the $N\times N$ force constant matrix, and $\eta$ is a
small positive number. $N$ is the degrees of freedom in the system.
$I$ is a unit matrix with the same dimension as $K$. As a result,
\begin{eqnarray}
G_{0}^{r}[0] & = & -K^{-1}.
\end{eqnarray}

The calculation of $G^{>}(0)$ is more complicated. However we find that it will be much more advantageous to work in the eigen space.

(a). Firstly, the force constant matrix $K$ is diagonalized:
\begin{eqnarray}
S^{\dagger}KS & = & \Omega^{2}
\end{eqnarray}
where the unitary matrix $S$ stores the information of all eigen vectors. And eigen values
are in the diagonalized matrix $\Omega^{2}$:
\begin{eqnarray}
\Omega^{2} & = & \left(\begin{array}{ccccc}
\omega_{0}^{2}\\
 & \ddots\\
 &  & \omega_{\mu}^{2}\\
 &  &  & \ddots\\
 &  &  &  & \omega_{N-1}^{2}\end{array}\right).
\end{eqnarray}

(b). The GF in real space (including both time domain and frequency domain) can be obtained from the corresponding one
in the eigen space. We take $G_{0}^{r}[\omega]$ as an example to illustrate this correspondence:
\begin{eqnarray}
G_{0}^{r}[\omega] & = & [(\omega+i\eta)^{2}I-K]^{-1}\nonumber\\\nonumber
 & = & [(\omega+i\eta)^{2}I-S\Omega^{2}S^{\dagger}]^{-1}\\\nonumber
 & = & S[(\omega+i\eta)^{2}I-\Omega^{2}]^{-1}S^{\dagger}\\
 & = & SG_{0}^{r}\{\omega\}S^{\dagger},
\end{eqnarray}
where we have used the relation for two matrices $(AB)^{-1}=B^{-1}A^{-1}$ and
we have introduced notation $G_{0}^{r}\{\omega\}=[(\omega+i\eta)^{2}I-\Omega^{2}]^{-1}$
to denote the GF in the eigen space. $G_{0}^{r}\{\omega\}$ is a diagonalized
matrix with the element as
\begin{eqnarray}
G_{0}^{r}\{\omega,\omega_{\mu}\} & = & [(\omega+i\eta)^{2}-\omega_{\mu}^{2}]^{-1}.
\end{eqnarray}
Similarly we have,
\begin{eqnarray}
G_{0}^{>}(t) =  S\left(G_{0}^{>}\{t\}\right)S^{\dagger}.
\label{eq_Ggt_t}
\end{eqnarray}
This kind of correspondence is valid for other GF, and the relations between different GF in the real space still hold in the eigen space.

(d). From Zeng's thesis\cite{Zeng} or P. Brouwer's note\cite{Brouwer}, the GF for single oscillator system
with frequency $\omega_{0}$ has explicit expression in time
domain and frequency domain:
\begin{eqnarray}
G_{0}^{r}(t) & = & \frac{-i}{2\omega_{0}}\theta(t)\left(e^{-i\omega_{0}t}-e^{i\omega_{0}t}\right),\\
G_{0}^{r}[\omega] & = & \frac{1}{(\omega+i\eta)^{2}-\omega_{0}^{2}},\\
G_{0}^{a}(t) & = & \frac{i}{2\omega_{0}}\theta(-t)\left(e^{-i\omega_{0}t}-e^{i\omega_{0}t}\right),\\
G_{0}^{a}[\omega] & = & \frac{1}{(\omega-i\eta)^{2}-\omega_{0}^{2}},\\
G_{0}^{<}(t) & = & \frac{-i}{2\omega_{0}}\left[(1+f)e^{i\omega_{0}t}+fe^{-i\omega_{0}t}\right],\\
G_{0}^{<}[\omega] & = & \frac{-i\pi}{\omega_{0}}\left[\delta(\omega+\omega_{0})(1+f)+\delta(\omega-\omega_{0})f\right],\\
\label{eq_Ggt_eigen}
G_{0}^{>}(t) & = & \frac{-i}{2\omega_{0}}\left[(1+f)e^{-i\omega_{0}t}+fe^{i\omega_{0}t}\right],\\
G_{0}^{>}[\omega] & = & \frac{-i\pi}{\omega_{0}}\left[\delta(\omega-\omega_{0})(1+f)+\delta(\omega+\omega_{0})f\right],
\end{eqnarray}
where $f=f(\omega_{0})=\frac{1}{e^{\beta\hbar\omega_{0}}-1}$ is the Bose
distribution function with $\beta=1/(k_{B}T)$.

(e). Using Eq.~(\ref{eq_Ggt_t}) and ~(\ref{eq_Ggt_eigen}), the one point GF $G_{j}$ can be calculated without doing any integration:
\begin{eqnarray}
G_{j} & = & \sum_{lmn}k_{lmn}G_{lm}^{>}(0)G_{nj}^{r}[0]\nonumber\\
 & = & \sum_{lmn}k_{lmn}\left(SG_{0}^{>}\{t=0\}S^{\dagger}\right)_{lm}\left(-K^{-1}\right)_{nj}\nonumber\\
 & = & (-i)\sum_{lmn}k_{lmn}\left(S\left(\begin{array}{ccc}
\ddots\nonumber\\
 & \frac{1}{2\omega_{\mu}}(2f+1)\nonumber\\
 &  & \ddots\end{array}\right)S^{\dagger}\right)_{lm}\nonumber\\
&&\times\left(-K^{-1}\right)_{nj}.\end{eqnarray}

In the study of the coefficient of thermal expansion, the derivative
of $G_{j}(t)$ with respective to temperature $T$ is needed:
\begin{eqnarray}
\frac{dG_{j}}{dT} & = & (-i)\sum_{lmn}k_{lmn}\left(S\left(\begin{array}{ccc}
\ddots\nonumber\\
 & \frac{1}{\omega_{\mu}}(\frac{df}{dT})\nonumber\\
 &  & \ddots\end{array}\right)S^{\dagger}\right)_{lm}\nonumber\\
&&\times\left(-K^{-1}\right)_{nj}.
\end{eqnarray}

\end{document}